\RequirePackage{filecontents}

\RequirePackage{snapshot}
\documentclass[aps,onecolumn,notitlepage,secnumarabic,superscriptaddress]{preprint}

\def\Preprint{}
\renewcommand\documentclass[2][]{}

\documentclass[a4paper,intlimits]{spie}

\usepackage[T1]{fontenc}
\usepackage{amsmath}
\usepackage{txfonts}
\usepackage{isomath}
\makeatletter
\renewcommand*{\vec}[1]{\ifcat a#1\mathbfit{#1}\else\boldsymbol{#1}\fi}
\makeatother
\usepackage{graphicx}
\graphicspath{{figures/}}
\usepackage[american]{babel}
\usepackage{microtype}
\usepackage{textcomp}
\usepackage{braket}

\newcommand*{\op}[1]{{\hat#1}}
\newcommand*{\I}{\mathrm{i}}
\newcommand*{\D}{\mathrm{d}}
\newcommand*{\e}{\mathrm{e}}
\newcommand*{\trans}[1]{#1^{\mathsf{T}}}%

\newcommand*{\acommut}[2]{%
  \mathchoice{\left\{#1,#2\right\}}{\{#1,#2\}}{\{#1,#2\}}{\{#1,#2\}}%
}
\newcommand*{\ft}[1]{\mathcal{F}\left\{#1\right\}}
\newcommand*{\ftinv}[1]{\mathcal{F}^{-1}\left\{#1\right\}}
 
\begin{document} 

\title{Computational relativistic quantum dynamics and its application
  to relativistic tunneling and Kapitza-Dirac scattering}  

\ifdefined\Preprint 
\author{Heiko~Bauke}
\email{heiko.bauke@mpi-hd.mpg.de}
\author{Michael~Klaiber}
\author{Enderalp~Yakaboylu}
\author{Karen~Z.~Hatsagortsyan}
\author{Sven~Ahrens}
\affiliation{Max-Planck-Institut f{\"u}r Kernphysik,
  Saupfercheckweg~1, 69117~Heidelberg, Germany}
\author{Carsten~M{\"u}ller}
\affiliation{Max-Planck-Institut f{\"u}r Kernphysik,
  Saupfercheckweg~1, 69117~Heidelberg, Germany}
\affiliation{Institut f{\"u}r Theoretische Physik~I,
  Heinrich-Heine-Universit{\"a}t D{\"u}sseldorf, Universit{\"a}tsstra{\ss}e~1,
  40225~D{\"u}sseldorf, Germany}
\author{Christoph~H.~Keitel}
\affiliation{Max-Planck-Institut f{\"u}r Kernphysik,
  Saupfercheckweg~1, 69117~Heidelberg, Germany}
\else
\author{Heiko~Bauke\supit{a}, 
  Michael~Klaiber\supit{a}, 
  Enderalp~Yakaboylu\supit{a}, 
  Karen~Z.~Hatsagortsyan\supit{a}, 
  Sven~Ahrens\supit{a}, Carsten~M{\"u}ller\supit{a,b}, 
  Christoph~H.~Keitel\supit{a}
  \skiplinehalf
  \supit{a}Max-Planck-Institut f{\"u}r Kernphysik,
  Saupfercheckweg~1,~69117~Heidelberg, Germany;\\
  \supit{b}Institut f{\"u}r Theoretische Physik~I,
  Heinrich-Heine-Universit{\"a}t D{\"u}sseldorf, Universit{\"a}tsstra{\ss}e~1,
  40225~D{\"u}sseldorf, Germany
}

\authorinfo{Corresponding author: Heiko Bauke, 
  e-mail: heiko.bauke@mpi-hd.mpg.de}

\maketitle 

\fi

\begin{abstract}
  Computational methods are indispensable to study the quantum dynamics
  of relativistic light-matter interactions in parameter regimes where
  analytical methods become inapplicable.  We present numerical methods
  for solving the time-dependent Dirac equation and the time-dependent
  Klein-Gordon equation and their implementation on high performance
  graphics cards.  These methods allow us to study tunneling from
  hydrogen-like highly charged ions in strong laser fields and
  Kapitza-Dirac scattering in the relativistic regime.
\end{abstract}


\ifdefined\Preprint 
\maketitle 
\else
\fi

\section{Introduction}
\label{sec:intro} 

With present-day strong lasers \cite{Yanovsky:etal:2008:Ultra-high}
providing intensities of the order of $10^{22}\,\mathrm{W/cm^2}$ and
employing highly charged ions, the regime of relativistic quantum
dynamics of strong field processes becomes accessible
\cite{DiPiazza:etal:2012:high-intensity}.  The time-dependent
Klein-Gordon equation and the time-dependent Dirac equation form the
basis for a theoretical description of the relativistic quantum dynamics
of spin-zero and spin-half particles, respectively.  Deducing analytical
solutions of these equations, however, poses a major problem.
Analytical methods for determining solutions of these equations usually
require physical setups with a high degree of symmetry
\cite{Bagrov:Gitman:1990:Exact_Solutions,Gross:1999:Relativistic_QM,Strange:1998:Relativistic_QM,Schwabl:2008:Advanced_QM,Wachter:2009:Relativistic_QM,Ehlotzky:etal:2009:Fundamental_processes}.
For the interaction with high-frequency few-cycle laser pulses of high
intensity also approximation methods reach their limits. Thus, ultra
short laser-matter interactions at relativistic intensities require
numerical approaches via computer simulations.

For this purpose, mature methods from nonrelativistic computational
quantum dynamics may be transferred into the relativistic regime.
However, also new and better methods are needed because relativistic
computational quantum dynamics is much more demanding than its
nonrelativistic sibling. In this contribution, we survey the challenges
of numerical time-dependent relativistic quantum dynamics and present
approaches to master these challenges by smart numerical algorithms
\cite{Ruf:2009:Klein-Gordon_Splitoperator,Blumenthal:Bauke:2012:stability_analysis},
by high-performance implementations on parallel architectures, including
cluster computers and graphics processing cards
\cite{Ruf:2009:Klein-Gordon_Splitoperator,Bauke:Keitel:2011:Accelerating,Bauke:Mertens:2005:ClusterComputing}.
Another approach is to cast a quantum system's mathematical description
by physical insights into a form that is beneficial for numerical
methods \cite{Bauke:Keitel:2009:Canonical_transforms}.  The new methods
of computational relativistic quantum dynamics are likely to be
beneficial for the emerging field of computational quantum
electrodynamics \cite{Wagner:etal:2011:Space-Time_Resolved}.

This contribution is organized as follows. In section~\ref{sec:cqd} we
will present numerical approaches to solve time-dependent relativistic
quantum equations of motion.  Later, some applications of numerical
relativistic time-dependent quantum dynamics will be highlighted. The
tunneling dynamics in relativistic strong-field ionization will be
investigated in section~\ref{sec:tunnel} by means of numerical (and
analytical) methods to develop an intuitive picture for the relativistic
tunneling regime \cite{Klaiber:etal:2012:Under-the-barrier_dynamics}.
It is demonstrated that the well-known tunneling picture applies also in
the relativistic regime by introducing position dependent energy
levels. The time for the formation of momentum components of the ionized
electron wave packet (Keldysh time) and the time interval which the
electron wave packet spends inside the barrier (Eisenbud-Wigner-Smith
time delay) are identified as the two characteristic time scales of
relativistic tunnel ionization. The Keldysh time can be related to a
momentum shift that is shown to be present in the ionization spectrum at
the detector and, therefore, observable experimentally. In
section~\ref{sec:KDE} we apply computational relativistic quantum
dynamics to study Kapitza-Dirac scattering of electrons by light
\cite{Ahrens:etal:2012:spin_dynamics}.  In particular, we demonstrate
that inelastic scattering processes involving a small odd number of
photons are intrinsically relativistic and involve periodic oscillations
of the electron's spin degree of freedom.

\section{Computational relativistic quantum dynamics}
\label{sec:cqd} 

Relativistic quantum dynamics in regimes where the single particle
picture is applicable is governed by the Klein-Gordon equation and the
Dirac equation.  The Klein-Gordon equation, a relativistic
generalization of the Schr{\"o}dinger equation, is an equation of motion
for a wave function $\Psi(\vec r, t)$
\cite{Feshbach:Villars:1954:Relativistic_Wave_Mechanics,Ruf:2009:Klein-Gordon_Splitoperator}.
It governs the evolution of a charged spinless particle with mass $m$
and charge $q$ moving under the effect of electrodynamic potentials
$\vec A(\vec r, t)$ and $\phi(\vec r, t)$.  Introducing the speed of
light $c$, the Klein-Gordon equation reads in its two-component
representation
\begin{equation}
  \label{eq:Klein-Gordon}
  \I\hbar\frac{\partial \Psi(\vec r, t)}{\partial t}
   = \op H_\mathrm{KG}(t)\Psi(\vec r, t) =
  \left(
    \frac{\sigma_3+\I\sigma_2}{2m}
    \left( -\I\hbar\nabla - q \vec A(\vec r, t) \right)^2
    + q \phi(\vec r, t) + \sigma_3 mc^2
  \right)\Psi(\vec r, t)\,.
\end{equation}
The wave function's components are related by the Pauli matrices
$\sigma_2$ and $\sigma_3$.  A possible representation reads
\begin{align}
  \label{eq:Pauli_matrices}
  \sigma_0 & = \begin{pmatrix} 1&0\\ 0&1 \end{pmatrix}\,{,} &
  \sigma_1 & = \begin{pmatrix} 0&1\\ 1&0 \end{pmatrix}\,{,} &
  \sigma_2 & = \begin{pmatrix} 0&-\I\\ \I&0 \end{pmatrix}\,{,} &
  \sigma_3 & = \begin{pmatrix} 1&0\\ 0&-1 \end{pmatrix} \,.
\end{align}
The physical electromagnetic fields follow from the potentials via
\begin{align}
  \vec E(\vec r, t) & = 
  -\nabla\phi(\vec r, t) - \frac{\partial\vec A(\vec r, t)}{\partial t} \,,\\
  \vec B(\vec r, t) & = 
  \nabla\times \vec A(\vec r, t) \,.
\end{align}
Similarly, the Dirac equation, a relativistic generalization of the
Pauli equation, is an equation of motion for a four-componet wave
function $\Psi(\vec r, t)$ that governs the evolution of a charged
spin-half particle.  It reads
\begin{equation}
  \label{eq:Dirac}
  \I\hbar\frac{\partial \Psi(\vec r, t)}{\partial t} =
  \op H_\mathrm{D}(t)\Psi(\vec r, t) = 
  \left(
    c
    \vec\alpha\cdot\left(
      -\I\hbar\nabla - q \vec A(\vec r, t)
    \right) + q \phi(\vec r, t) + mc^2\beta
  \right)\Psi(\vec r, t)
\end{equation}
with the matrices $\vec\alpha=\trans{(\alpha_1, \alpha_2, \alpha_3)}$
and $\beta$ obeying the algebra
\begin{equation}
  \label{eq:Dirac_algebra}
  \alpha_i^2 = \beta^2 = 1\,,\quad
  \alpha_i\alpha_k + \alpha_k\alpha_i = 2\delta_{i,k}\,,\quad
  \alpha_i\beta + \beta\alpha_i = 0\,.
\end{equation}
This algebra determines the matrices $\alpha_i$ and $\beta$ only up to
unitary transforms. In numerical applications, we adopted the so-called
Dirac representation with
\begin{equation}
  \label{eq:Dirac_repr}
  \begin{aligned}
    \alpha_1 & =
    \begin{pmatrix}
      0 & \sigma_1 \\
      \sigma_1 & 0
    \end{pmatrix}\,, &
    \alpha_2 & =
    \begin{pmatrix}
      0 & \sigma_2 \\
      \sigma_2 & 0
    \end{pmatrix}\,, &
    \alpha_3 & =
    \begin{pmatrix}
      0 & \sigma_3 \\
      \sigma_3 & 0
    \end{pmatrix}\,, &
    \beta & =
    \begin{pmatrix}
      \sigma_0 & 0 \\
      0 & -\sigma_0
    \end{pmatrix}\,.
  \end{aligned}
\end{equation}

While the numerical solution of the time-dependent Schr{\"o}dinger equation
is a mature field with many applications in physics and chemistry
relativistic computational quantum dynamics has not yet reached the same
degree of maturity.  This is partly due to the fact that the interest in
solving the relativistic quantum dynamics
\cite{Joachain:1997:1D_Dirac,Rathe:elal:1997:Intense_laser-atom,Taieb:etal:1998:Signature,Braun:etal:1999:Numerical_approach,Mocken:Keitel:2004:Quantum_dynamics,Mocken:Keitel:2008:FFT-split-operator,Selsto:etal:2009:Solution_of_the_Dirac_equation,Fillion-Gourdeau:etal:2012:time-dependent_Dirac,Vanne:Saenz:2012:multiphoton_ionization}
has grown as recently as high-intensity laser facilities where proposed
for the near future or even became available providing access to
experiments for quantum dynamics in the relativistic regime.
Furthermore, attempts to solve the Klein-Gordon or Dirac equation were
also hampered by the high computational resources that are required.
Numerical procedures propagate quantum wave functions in discrete time
steps.  The smaller the time step the larger the computational cost.
The maximal time step $\upDelta t$ that a numerical procedure may take
is limited by the typical energy $\tilde{E}$ of the system under
consideration \cite{Bauke:Keitel:2009:Canonical_transforms},
viz. $\upDelta t\leq\pi\hbar/|\tilde{E}|$.  In relativistic quantum
dynamics, we have typical energies of the order $\pm mc^2$ whereas for
the nonrelativistic dynamics the restmass energy is irrelevant.
Assuming the ground state energy of a hydrogen atom and the restmass
energy of an electron as typical energy scales of nonrelativistic and
relativistic quantum dynamics we find $\upDelta t\leq 48\,\mathrm{as}$
and $\upDelta t\leq 0.0013\,\mathrm{as}$, respectively.  Thus, solving
the time-dependent Klein-Gordon or Dirac equation for a given time
interval requires a number of time steps that is several orders of
magnitude larger than for the nonrelativistic Schr{\"o}dinger equation.

A partial solution to this problem is provided by the Foldy-Wouthuysen
transformation \cite{Foldy:Wouthuysen:1950:Non-Relativistic_Limit} which
transforms the Dirac equation (\ref{eq:Dirac}) into block diagonal form.
In leading order the new Hamiltonian reads with
$\vec\sigma=\trans{(\sigma_1, \sigma_2, \sigma_3)}$
\begin{multline}
  \label{eq:FW}
  \op H_\mathrm{D}'(t) =
  \beta\left(
    mc^2 + \frac{1}{2m}\left(
      -\I\hbar\nabla-q\vec{A}(\vec r, t)
    \right)^2 - 
    \frac{q\hbar}{2m}\vec\sigma\cdot\vec{B}(\vec r, t)
  \right) + q\phi(\vec r, t) \\
  -\beta\left(
    \frac{1}{8m^3c^2}\left(
      -\I\hbar\nabla-q\vec{A}(\vec r, t)
    \right)^4 +
    \frac{q^2\hbar^2}{8m^3c^2}\vec{B}(\vec r, t)^2 -
    \frac{q\hbar}{8m^3c^2}\acommut{\vec\sigma\cdot\vec{B}(\vec r, t)}{\left(
        -\I\hbar\nabla-q\vec{A}(\vec r, t)
      \right)^2} 
  \right) \\
  -
  \frac{q\hbar^2}{8m^2c^2}\nabla\cdot\vec{E}(\vec r, t) -
  \frac{\I q\hbar^2}{8m^2c^2}\vec\sigma\cdot(\nabla\times\vec{E}(\vec r, t)) -
  \frac{q\hbar}{4m^2c^2}\vec\sigma\cdot(\vec{E}(\vec r, t)\times(-\I\hbar\nabla))
  + \dots
\end{multline}
In this representation, the Dirac equation decouples into two equations
for two two-component wave functions having typical energies of order
$+mc^2$ and $-mc^2$, respectively. The additive restmass terms $\pm
mc^2$ in (\ref{eq:FW}) may be removed by a gauge transform allowing for
an efficient numerical solution \cite{Hu:Keitel:2001:Dynamics}.
However, the representation (\ref{eq:FW}) is of limited practical value.
Compared to the original Dirac equation (\ref{eq:Dirac}) the Hamiltonian
(\ref{eq:FW}) is rather complicated although it represents only the
first terms of an infinite series. It is applicable in the weakly
relativistic limit only. Exact Foldy-Wouthuysen transformations are
known for a certain classes of potentials only
\cite{Nikitin:1998:exact_Foldy_Wouthuysen}.  Therefore, exact numerical
solutions of the time-dependent Klein-Gordon or Dirac equation require
high performance computer implementations of numerical propagation
schemes or taking advantage of special features of the physical system
under consideration.

Aiming for high performance, computer implementations of numerical
propagation schemes utilize parallel computing architectures.
Currently, there are two major directions of development, multicore
architectures with a few (typically two to a few dozen in the near
future) general purpose computing units and dedicated accelerators,
e.\,g., graphics processing units (GPUs), with several hundred or more
computing units with reduced capabilities each as compared to
traditional central processing units (CPUs)
\cite{Kirk:Hwu:2010:Programming_MPP,Kindratenko:etal:2010:_HPC_Accelerators}.
Employing GPUs to perform computations that are traditionally handled by
CPUs is commonly referred to as GPU computing
\cite{Owens:etal:2008:GPU_Computing}.

Using Dyson's time ordering operator $\hat T$ the formal solution
\cite{Pechukas:Light:1966:Time-Displacement} to the time-dependent
Klein-Gordon equation and the time-dependent Dirac equation reads
\begin{equation}
  \label{eq:prop}
  \Psi(\vec r, t+\Delta t) = 
  \op T\exp\left(
    -\frac{\I}{\hbar}\int_{t}^{t+\Delta t} \op H(t') \,\D t' 
  \right)\Psi(\vec r, t) \,.
\end{equation}
Usually it is not possible to cast (\ref{eq:prop}) directly into a
numerical scheme. Some approximations have to be introduced.  Splitting
the Hamiltonian $\op H(t)$ into $\op H(t)=\op H_1(t)+\op H_2(t)$ the
expression (\ref{eq:prop}) may be expanded into
\cite{Feit:etal:1982:spectral_method}
\begin{equation}
  \Psi(\vec r, t+\Delta t) =  
  \underbrace{
    \exp\left(
      -\frac{\I}{2\hbar}\int_{t}^{t+\Delta t} \op H_{1}(t') \,\D t' 
    \right)}_{\displaystyle=\op U_{\op H_{1}}(t+\Delta t, t)}
  \underbrace{
    \exp\left(
      -\frac{\I}{\hbar}\int_{t}^{t+\Delta t} \op H_{2}(t') \,\D t' 
    \right)}_{\displaystyle=\op U_{\op H_{2}}(t+\Delta t, t)} 
  \underbrace{
    \exp\left(
      -\frac{\I}{2\hbar}\int_{t}^{t+\Delta t} \op H_{1}(t') \,\D t' 
    \right)}_{\displaystyle=\op U_{\op H_{1}}(t+\Delta t, t)}
  \Psi(\vec r, t) + O\left(\Delta t^3\right)
\end{equation}
leading to the (Fourier) split operator method
\cite{Feit:etal:1982:spectral_method,Mocken:Keitel:2008:FFT-split-operator,Ruf:2009:Klein-Gordon_Splitoperator,Bauke:Keitel:2011:Accelerating}.
In the case of the Dirac equation we split the Dirac Hamiltonian
(\ref{eq:Dirac}) into
\begin{subequations}
  \begin{align}
    \op H_{\mathrm{D}, 1}(t) & =
    -cq \vec\alpha\cdot\vec{A}(\vec r, t) + q\phi(\vec r, t) \,,\\
    \op H_{\mathrm{D}, 2}(t) & = -\I c\hbar\vec\alpha\cdot\nabla +
    mc^2\beta \,.
  \end{align}
\end{subequations}%
\begin{figure}
  \centering
  \includegraphics[scale=0.5]{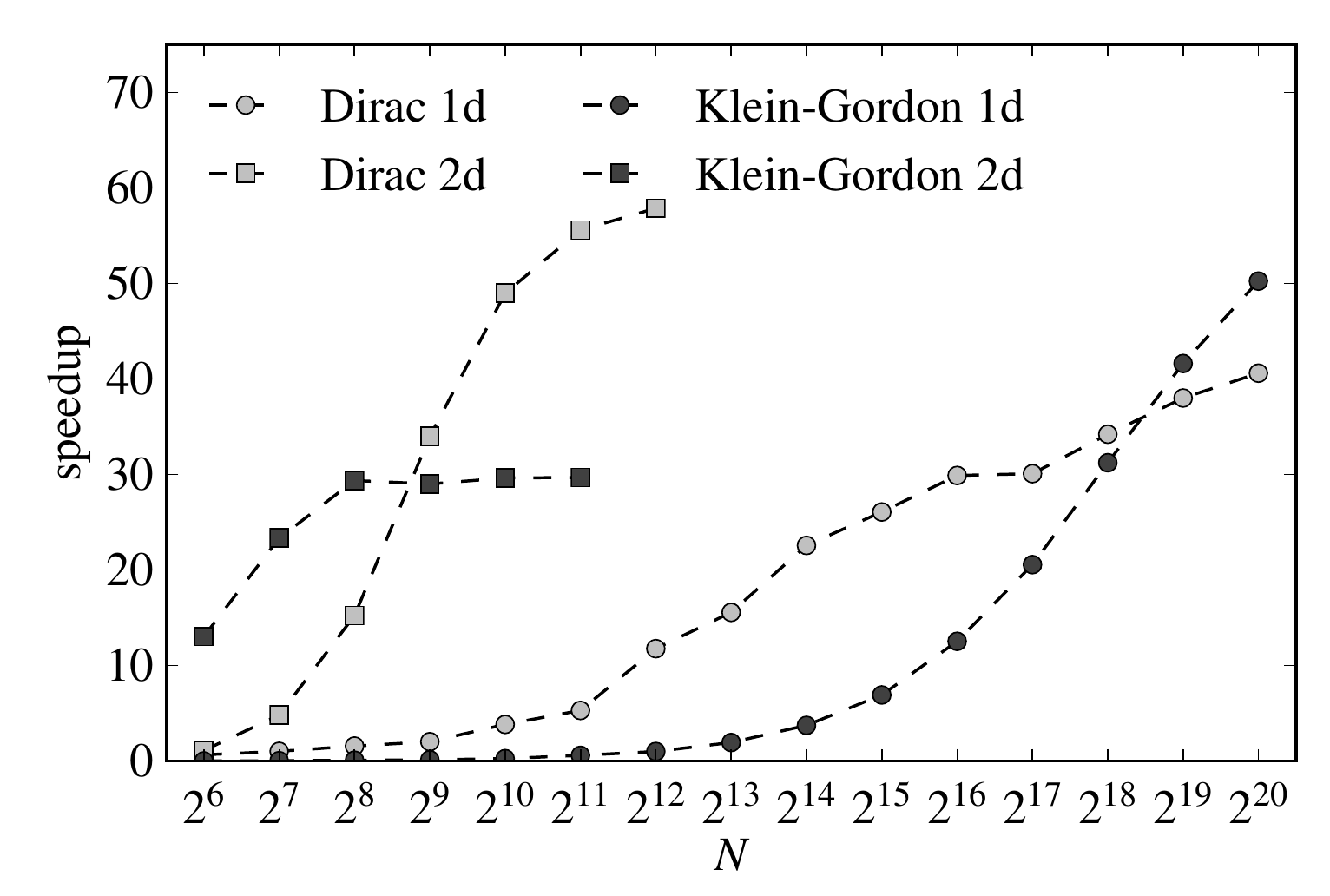}
  \caption{Speedup for propagating a Dirac wave function and a
    Klein-Gordon wave function over 128 time steps as a function of the
    grid size. Circles depict results for the one-dimensional
    Klein-Gordon and Dirac equation on a grid of $N$ data points while
    squares shows results for the two-dimensional case on a grid of
    $N\times N$ data points. Performance measurements carried out on a
    GeForce~GTX~480 GPU utilizing the CUFFT library and double precision
    floating point numbers. A single CPU core (Intel Core i7 CPU at
    2.93\,GHz) and the FFTW3 library serves as a reference.}
  \label{fig:speedup}
\end{figure}%
The operator $\op U_{\op H_{\mathrm{D},1}}(t+\Delta t, t)$ is diagonal
in real space and $\op U_{\op H_{\mathrm{D},2}}(t+\Delta t, t)$ is
diagonal in Fourier space. Thus, the Fourier split operator method
propagates $\Psi(\vec r, t)$ alternating in real and in Fourier space,
viz.
\begin{equation*}
  \Psi(\vec r, t+\Delta t) = 
  \op U_{\op{H}_{\mathrm{D},1}}\left(t + \Delta t, t\right)
  \ftinv{
    \op{{\tilde U}}_{\op{H}_{\mathrm{D},2}}\left(t + \Delta t, t\right)
    \ft{\op U_{\op{H}_{\mathrm{D},1}}\left(t + \Delta t, t\right)\Psi(\vec r, t)}
  } 
  + O\left(\Delta t^3\right) \,,
\end{equation*}
where $\op{{\tilde U}}_{\op{H}_{\mathrm{D},2}}\left(t + \Delta t,
  t\right)$ denotes the Fourier space representation of
$\op{U}_{\op{H}_{\mathrm{D},2}}\left(t + \Delta t, t\right)$ and
$\ft{\cdot}$ and $\ftinv{\cdot}$ represent the Fourier transform and its
inverse, respectively.  From a computational point of view calculating
the Fourier transform and its inverse are the main tasks of the Fourier
split operator method.  GPUs provide an economically priced hardware
platform that allows very efficient implementations of the fast Fourier
transform leading to speedup factors of about 40 compared to singe CPU
implementations for solving the time-dependent Dirac equation, see
figure~\ref{fig:speedup}.  In the split operator approach to the
Klein-Gordon equation we partition (\ref{eq:Klein-Gordon}) into
\begin{subequations}
  \begin{align}
    \op H_{\mathrm{KG}, 1}(t) & = 
    q \phi(\vec r, t) + \sigma_3 mc^2 \,,\\
    \op H_{\mathrm{KG}, 2}(t) & =
    \frac{\sigma_3+\I\sigma_2}{2m}
    \left( -\I\hbar\nabla - q \vec A(\vec r, t) \right)^2 \,.
  \end{align}
\end{subequations}
In contrast to the Dirac equation, both operators $\op U_{\op
  H_{\mathrm{KG},1}}(t+\Delta t, t)$ and $\op U_{\op
  H_{\mathrm{KG},2}}(t+\Delta t, t)$ can be computed efficiently in real
space.  Employing a finite difference representation of the operators
$\op H_{\mathrm{KG}, 1}(t)$ and $\op H_{\mathrm{KG}, 2}(t)$ allows
parallel numerical algorithms suitable for distributed memory computers
as well as for GPUs.  See figure~\ref{fig:speedup} for benchmarks on the
real space split operator method for the Klein-Gordon equation.  Note
that the real space split operator method for the Klein-Gordon equation
reaches on a low-cost consumer graphics card a similar performance as on
a cluster computer with a few dozen CPU cores
\cite{Ruf:2009:Klein-Gordon_Splitoperator}.

A complementary approach of solving relativistic quantum wave equations
is taking advantage of special features of the physical problem under
consideration.  As an example let us consider a system in the presence
of a sinusoidal vector potential
\begin{equation}
  \vec{A}(\vec r, t) = \cos(\vec r\cdot\vec k)\vec{A}(t)\,.
\end{equation}
In momentum space, this potential couples modes only which are
$\hbar\vec k$ apart.  Thus, it is legitimate to expand the wave function
into a superposition of momentum eigenstates
\cite{Thaller:2004:Advanced_Visual_QM} $u^{\uparrow\downarrow,\pm}_{\vec
  p}\e^{\I\vec p\cdot \vec r/\hbar}$ of momentum $\vec p$ which are
simultaneous energy eigenstates of the free Dirac Hamiltonian with
positive ($+$) or negative ($-$) energy and spin eigenstates with
positive ($\uparrow$) or negative ($\downarrow$) spin, viz.
\begin{equation}
  \label{eq:series}
  \Psi(\vec r, t) = \sum_{n} \left(
    c_{n}^{\uparrow,+}(t) u^{\uparrow,+}_{\vec p + n\hbar\vec k} +
    c_{n}^{\downarrow,+}(t) u^{\downarrow,+}_{\vec p + n\hbar\vec k} +
    c_{n}^{\uparrow,-}(t) u^{\uparrow,-}_{\vec p + n\hbar\vec k} +
    c_{n}^{\downarrow,-}(t) u^{\downarrow,-}_{\vec p + n\hbar\vec k} 
  \right)\e^{\I(\vec p+n\hbar\vec k)\cdot \vec r/\hbar}\,.
\end{equation}
Plugging the series (\ref{eq:series}) into the Dirac equation
(\ref{eq:Dirac}) yields an infinite system of ordinary linear
differential equations for the expansion coefficients
$c_{n}^{\uparrow\downarrow,\pm}(t)$.  The system's coupling matrix has a
banded structure.  Thus, the time evolution of the expansion
coefficients can be calculated efficiently by employing a Crank-Nicolson
scheme or Krylov subspace methods.  This approach is applicable in
particular to stetups with physical states that can be described by a
superposition of a small number of momentum eigenstates as in the
Kapitza-Dirac effect that we will study in section~\ref{sec:KDE}.

\section{Tunnel ionization at relativistic laser intensities}
\label{sec:tunnel} 

The ionization of hydrogen-like ions occurs in the so-called tunneling
regime when the laser's electric field amplitude $E_0$ is smaller than
those required for over-the-barrier ionization
\cite{Augst:etal:1989:Tunneling_ionization} and the Keldysh parameter
$\gamma=\omega\sqrt{2mI_p}/(eE_0)$ is well below unity with the
ionization potential $I_p$, the laser's angular frequency $\omega$ and
the elementary charge $e$. In the tunnel-ionization regime the electron
travels during its journey from the bound state into the continuum
through a region which is forbidden according to classical
mechanics. For nonrelativistic dynamics, when the typical electron
velocities are small compared to the speed of light, the effect of the
laser's magnetic field can be neglected. In this case the well-known
intuitive picture for the tunneling dynamics arises in which the
electron tunnels out through an effective potential $V_\mathrm{eff}(\vec
r) = V(\vec r) + \vec r \cdot \vec E(t)$ formed by the atomic potential
$V(\vec r)$ and the laser's oscillating electric field $\vec E(t)$.

However, the contribution of the laser's magnetic field component can no
longer be neglected when we enter the relativistic regime.  A transverse
laser field with perpendicular electric and magnetic components can not
be described solely by a scalar potential, a vector potential is
required which does not appear in the conventional tunneling picture.
Consequently, a generalization of the tunneling picture into the
relativistic regime is not straightforward and there is no clear
intuitive picture for the relativistic quasi-static ionization dynamics
in the classically forbidden region.  In the following we will establish
a tunneling picture that incorporates the vector potential by
quasi-classical considerations.  Predictions by this quasi-classical
picture will be verified by a numerical solution of the time-dependend
Dirac equation.

\begin{figure}[b]
  \centering
  \includegraphics[scale=0.5]{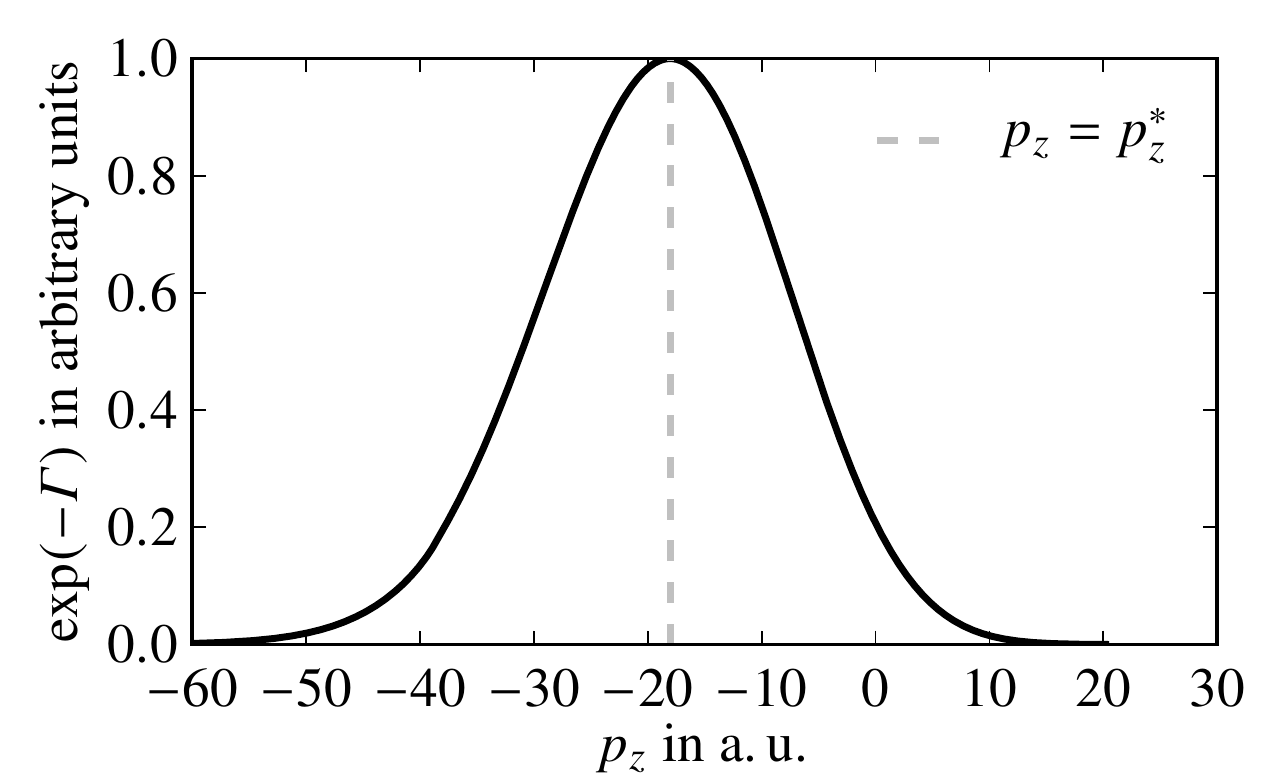}%
  \qquad
  \includegraphics[scale=0.5]{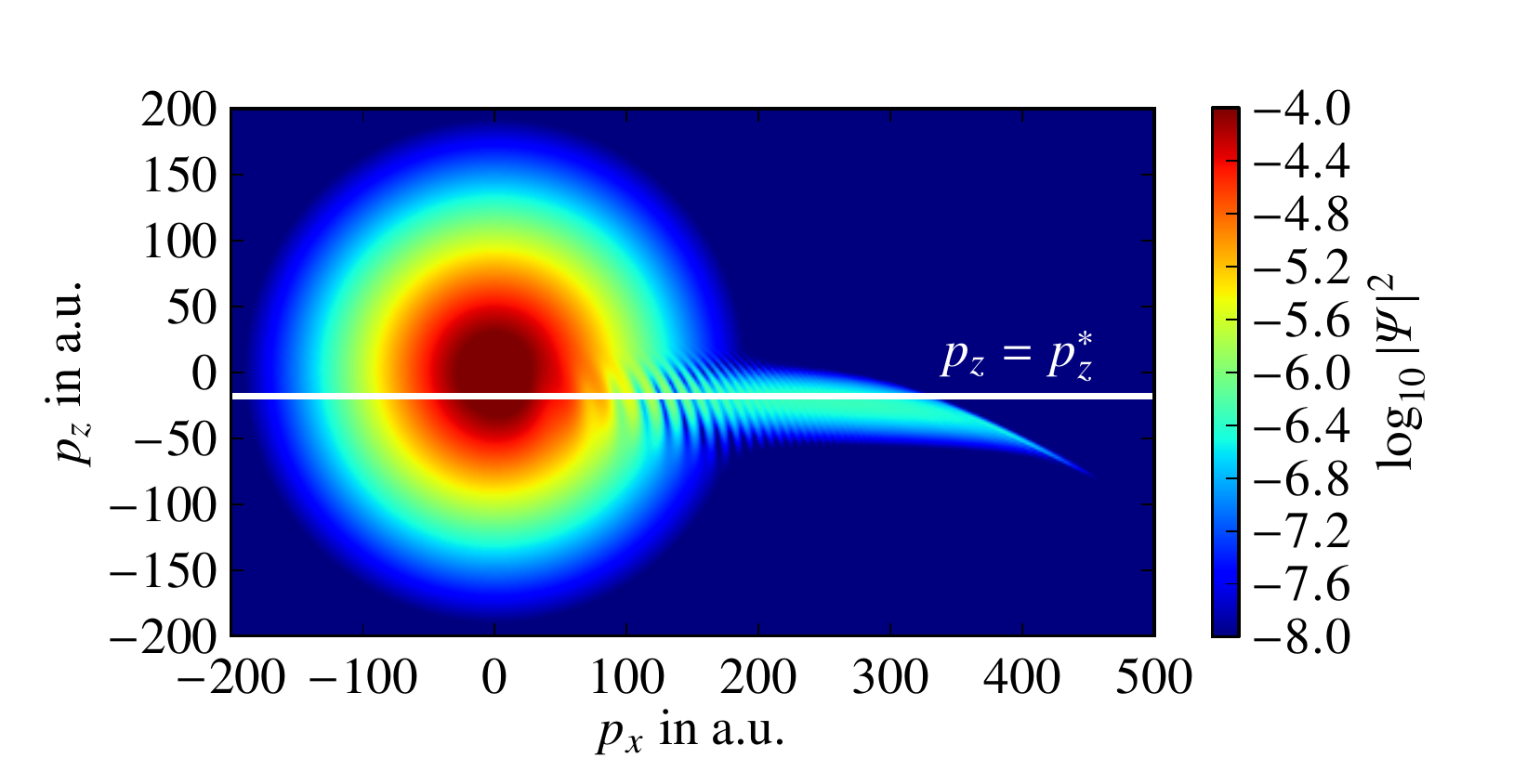}%
  \caption{Left panel: WKB tunneling probability for an electron in a
    two-dimensional soft-core potential (serving as a model for highly
    charged hydrogen-like ions) exposed to a strong laser field as a
    function of the canonical momentum in laser propagation direction
    $p_z$.  The tunneling probability is maximal at $p_z=p_z^*$. Right
    panel: The momentum space distribution of the corresponding wave
    function.  The density's portion at $p_z\approx p_z^*$ represents
    the tunneled part of the electronic wave function.  The applied
    parameters are $I_p/(mc^2)=1/4$ and $E_0/E_a=1/30$ with
    $E_a=5.14\cdot 10^{11}\,\mathrm{V/m}$ denoting the characteristic
    atomic field strength.}
  \label{fig:WKB}
\end{figure}

We describe the laser field via its vector potential in electric field
gauge \cite{Reiss:1979:gauge,Grynberg:2010:Quantum_Optics}, i.\,e.,
$\phi(\vec{r}, t)=-\vec{r}\cdot \vec{E}(\eta)$ and $\vec{A}(\vec{r},
t)=-\hat{\vec{k}}[\vec{r}\cdot \vec{E}(\eta)]/c$ with the laser phase
$\eta=t-z/c$, the electron's coordinate $\vec{r}=(x,y,z)$, the unit
vector in laser propagation direction $\hat{\vec{k}}$, and the speed of
light $c$.  In order to study the effect of the vector potential we
consider the static Hamiltonian at maximal laser field strength
\begin{equation}
  \label{eq:Hamiltonian}
  H = \frac{p_x^2}{2m} + \frac{p_y^2}{2m} + \frac{(p_z-qxE_0/c)^2}{2m} + 
  V(r)+qxE_0 \,,
\end{equation}
with the electron's charge $q=-e$, the canonical momenta $p_x$, $p_y$,
and $p_z$ and $r=\sqrt{x^2 + y^2 + z^2}$.  By virtue of the electric
field gauge the Hamilton function equals the total energy $-I_p$.  The
tunneling dynamics takes place in close vicinity along the
\mbox{$x$-axis} (the polarization direction). Therefore, the
atomic-potential's central force in the direction of the \mbox{$y$-axis}
and the \mbox{$z$-axis} (the laser's propagation directions) is
negligible in the tunneling region and the canonical momenta $p_y$ and
$p_z$ become approximately conserved.  Consequently, the electron
dynamics is separable and the equation $H=-I_p$ can be written as
\begin{equation} 
  \label{eq:1d}
  -I_p-\left(
    \frac{p_y^2}{2m} + \frac{(p_z-qxE_0/c)^2}{2m} 
  \right) = 
    \frac{p_x^2}{2m} + \left(V(x)+qxE_0\right)\,.
\end{equation}
Equation~(\ref{eq:1d}) is the desired generalization of the
nonrelativistic one-dimensional model of tunnel ionization
\cite{Augst:etal:1989:Tunneling_ionization}.  According to this
generalized model, the electron propagates along the laser polarization
direction through the barrier $V_\text{eff}(x)=V(x)+qxE_0$ with a
position dependent total energy
\begin{equation}
  \mathcal{E}(x)= - I_p-\left(
    \frac{p_y^2}{2m}+\frac{(p_z-qxE_0/c)^2}{2m}
  \right) \,,
\end{equation}
which is the difference between the binding energy and the kinetic
energy of the motion in the propagation direction of the laser.  

With (\ref{eq:1d}) we find the momentum in electric field direction
$p_x(x)=\I\sqrt{2m[V_\text{eff}(x)-\mathcal{E}(x)]}$.  The WKB tunneling
probability is proportional to $\exp(-\Gamma)$, where $\Gamma$ is given
by the integral
\begin{equation}
  \Gamma=-\frac{2\I}\hbar\int_{x_i}^{x_e} p_x(x)\,\mathrm{d}x
\end{equation}
over the classical forbidden region $x_i\le x \le x_e$, i.\,e., where
$V_\text{eff}(x)>\mathcal{E}(x)$ .  Note that the effective potential
depends on the canonical momenta in $y$- and $z$-direction.  The WKB
tunneling probability is maximal for $p_y=0$ and $p_z=p_z^*$.  Thus, the
most probable kinematic momenta with witch the electron enters and
leaves the barrier are $p_{z,\mathrm{kin}}(x_i)=p_z^*-qx_iE_0/c$ and
$p_{z,\mathrm{kin}}(x_e)=p_z^*-qx_eE_0/c$, respectively.
Figure~\ref{fig:WKB} (left panel) displays the WKB tunneling probability
as a function of the canonical momentum $p_z$.  The WKB tunneling
probability is maximal for $p_z^*\approx -2I_P/(3c)$ yielding a
kinematic momentum $p_{z,\mathrm{kin}}\approx I_p/(3c)$ at the tunneling
exit.  Thus, the quasi-classical model predicts that electrons leave the
tunneling barrier with non-zero momentum in propagation direction of the
laser as a consequence of the laser's magnetic field component. This may
be interpreted as a momentum transfer due to the Lorentz force acting on
the Keldysh-time scale $\tau_K$ during tunneling.  Numerical simulations
support the prediction of a momentum transfer.  Figure~\ref{fig:WKB}
(right panel) shows the distribution of the canonical momentum at
maximal laser intensity.  The momentum distribution consits of the bound
part which is symmetric around zero momentum and the ionized part at
about $p_z=p_z^*$.  Note that in contrast to the quasi-classical model
our numerical simulation includes all relativistic effects not only
magnetic dipole effects.  In fact, one can show that relativistic
effects that go beyond magnetic dipole effects change the overall
tunneling probability but do not affect the relativistic tunneling
picture as outlined above
\cite{Klaiber:etal:2012:Under-the-barrier_dynamics}.  Examining the wave
function in position space we find by a numerical solution of the Dirac
equation and quantum mechanical analytical calculations that the wave
function is also shifted during tunneling under the barrier into
propagation direction of the laser by about $\tau_E
p_{z,\mathrm{kin}}(x_e)$ with the Eisenbud-Wigner-Smith time delay
$\tau_E$ \cite{Klaiber:etal:2012:Under-the-barrier_dynamics}, i.\,e.,
the time the electron spends under the barrier.

\section{Kapitza-Dirac scattering at relativistic laser intensities}
\label{sec:KDE} 

The diffraction of an electron beam from a standing wave of light is
referred to as the Kapitza-Dirac effect
\cite{Kapitza:Dirac:1933:effect,Batelaan:2007:Illuminating_Kapitza-Dirac}.
It highlights the electron's quantum wave nature and may be interpreted
as an analogue of the optical diffraction of light on a massive grating,
but with the roles of light and matter interchanged.  In order to study
Kapitza-Dirac scattering at relativistic laser intensities we consider a
quantum electron wave packet in a standing linearly polarized light wave
with maximal electric field strength $\vec E$, wave vector $\vec k$, and
wavelength $\lambda=2\pi/|\vec k|=2\pi/k$, respectively. The laser is
modeled by the vector potential
\begin{equation}
  \vec A(\vec r,t) = 
  -\frac{\vec E}{ck} \cos(\vec k \cdot \vec r) \sin(c k t) w(t)
  \label{eq:A}
\end{equation}
introducing the speed of light $c$ and the temporal envelope function
$w(t)$.  The relativistic quantum dynamics of an electron with mass $m$
and charge $-e$ is governed by the Dirac equation (\ref{eq:Dirac}).
Employing the ansatz (\ref{eq:series}) and the potential (\ref{eq:A})
the Dirac equation yields a system of ordinary differential equations
\begin{multline}
  \I\hbar\dot c_n^\gamma(t) = 
  \varepsilon^\gamma \mathcal{E}(\vec p + n\hbar\vec k) c_n^\gamma(t) \\
  - 
  \frac{w(t) e \sin(c k t)}{2 k} \sum_{\zeta} \left(
    \braket{u_{\vec p+n\hbar\vec k}^\gamma|
      \vec E\cdot \vec \alpha|
      u_{\vec p+(n-1)\hbar\vec k}^{\zeta}} c_{n-1}^{\zeta}(t) +
    \braket{u_{\vec p+n\hbar\vec k}^\gamma|
      \vec E\cdot \vec \alpha|
      u_{\vec p+(n+1)\hbar\vec k}^{\zeta}} 
    c_{n+1}^{\zeta}(t) 
  \right)\,,
  \label{eq:FSME-dirac-equation}
\end{multline}
where we have introduced the relativistic energy momentum dispersion
relation $\mathcal{E}(\vec p)=\sqrt{m^2 c^4 + c^2 {\vec p}^2}$ and the
signum $\varepsilon^\gamma$, which is $1$ for $\gamma \in \{\uparrow{+},
\downarrow{+}\}$ and ${-}1$ for $\gamma \in \{\uparrow{-},
\downarrow{-}\}$.  The electron that is incident to the laser is modeled
by a positive-energy plane wave with momentum $\vec p$ and definite
spin. This means the initial condition for
(\ref{eq:FSME-dirac-equation}) is
\begin{equation}
  c_n^\gamma(0) =
  \begin{cases}
    1 & \text{if $n=0$ and $\gamma=\uparrow+$,} \\
    0 & \text{else.}
  \end{cases}
  \label{eq:FSME-dirac-equation_ini}
\end{equation}

\begin{figure}
  \centering
  \includegraphics[scale=0.5]{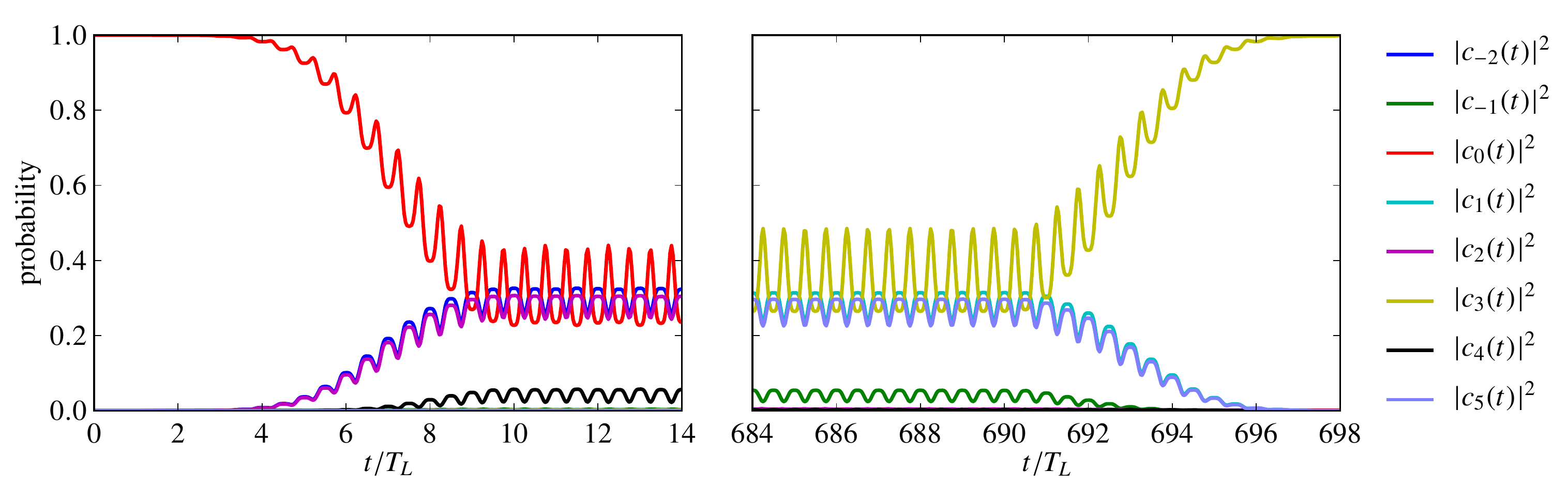}%
  \caption{Occupation of different momentum states
    $|c_n(t)|^2=|c_{n}^{\uparrow+}(t)|^2+|c_{n}^{\downarrow+}(t)|^2+
    |c_{n}^{\uparrow-}(t)|^2+|c_{n}^{\downarrow-}(t)|^2$ as a function
    of time $t$ measured in units of the lasers period $T_L$.  Laser
    parameters of this simulation correspond to two counterpropagating
    \mbox{X-ray} laser beams with a peak intensity of
    $2.0\times10^{23}\textrm{W}/\textrm{cm}^2$ each and a photon energy
    of $3.1\,\textrm{keV}$. The electron impinges at an angle of
    inclination of $\vartheta=0.4$\textdegree\ and a momentum of
    $176\,\textrm{keV}/c$, in order to fulfill the Bragg
    condition~(\ref{eq:gen_Bragg}).}
  \label{fig:KDE}
\end{figure}
\begin{figure}
  \centering
  \includegraphics[scale=0.5]{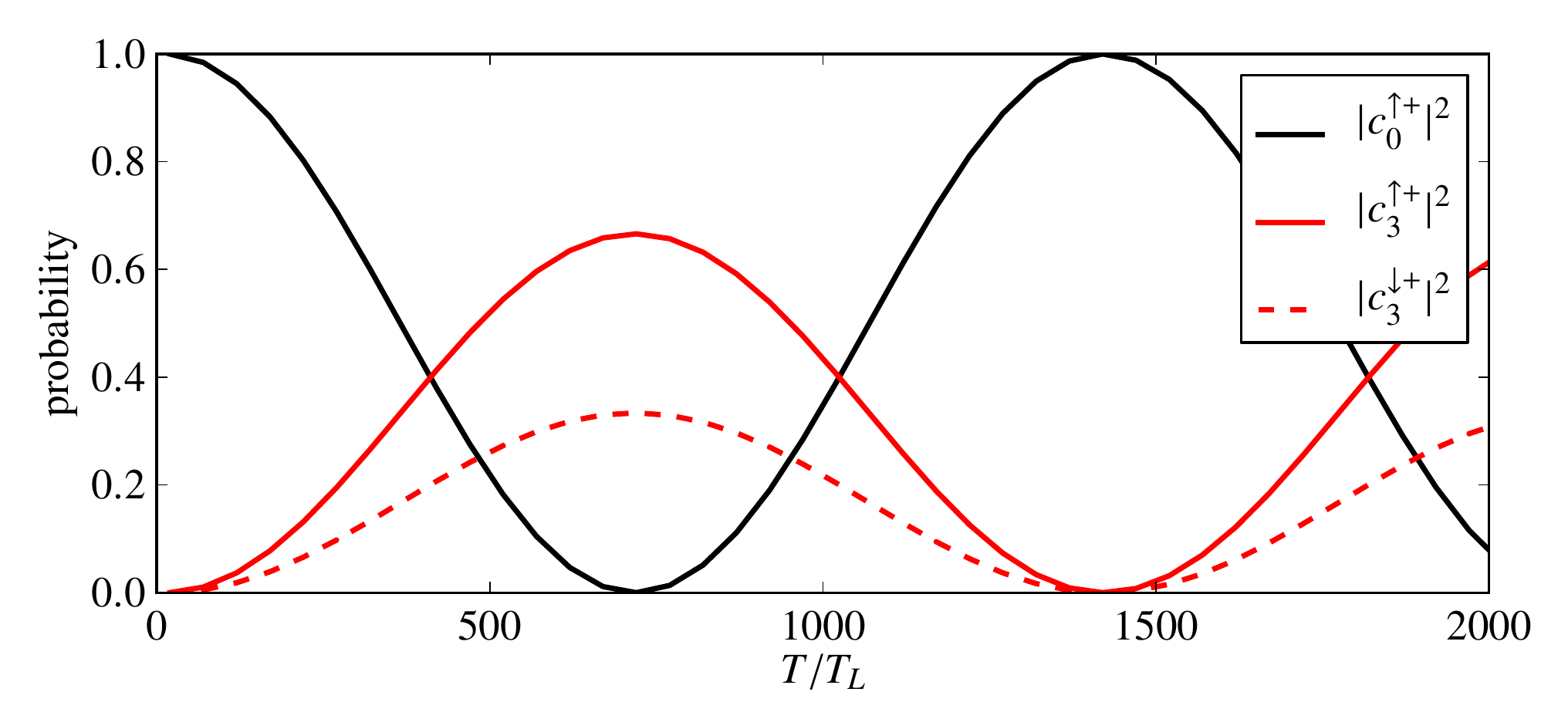}%
  \caption{Spin resolved occupation probabilities as a function of the
    interaction time $T$ for a \mbox{three-photon} Kapitza-Dirac effect
    (adopted from Phys. Rev. Let., vol.~109, article 043601, 2012).
    Experimental parameters as in figure~\ref{fig:KDE}.  The occupation
    probabilities oscillate in Rabi cycles with a period that is about
    1500 times larger than the laser period.}
  \label{fig:flip_0_3}
\end{figure}

Electron diffraction by light may be observed only if the electron's
momentum and the photon momenta of the laser meet a Bragg condition.
Considering classical energy and momentum conservation
\cite{Ahrens:etal:2012:spin_dynamics} yields
\begin{equation} 
  \label{eq:gen_Bragg}
  \frac{\cos\vartheta}{\lambda_p} = 
  -\frac{n_r-n_l}{2\lambda} + 
  \frac{n_r-n_l}{|n_r-n_l|}
  \frac{n_r+n_l}2\sqrt{
    \frac{1}{\lambda^2} -
    \frac{1}{n_r n_l}\left(
      \frac{\sin^2\vartheta}{\lambda^2_p} +
      \frac{1}{\lambda_C^2}
    \right)
  }
\end{equation}
with $\vartheta$ denoting the angle between $\vec p$ and $\vec k$, the
de~Broglie wavelength $\lambda_p=2\pi\hbar/|\vec p|$, and the Compton
wavelength $\lambda_C=2\pi\hbar/(mc)$.  The integers $n_r$ and $n_l$ are
the net numbers of photons exchanged with the right- and left-traveling
laser waves, respectively, with positive (negative) values indicating
photon absorption (emission).  To be consistent with the nonrelativistic
limit $n_r$ and $n_l$ must have opposite signs.  As a consequence of
(\ref{eq:gen_Bragg}) the initial electron momentum $\vec p$ and/or the
laser's photon momentum $\hbar k$ must be of the order of $m c$, i.\,e.,
relativistic, unless the electron scatters elastically, i.\,e.,
$n_r+n_l=0$, or the electron interacts with a very large number of
photons, i.\,e., $|n_r|+|n_l|\gg0$.

Figure~\ref{fig:KDE} shows a typical time evolution of the occupation of
different momentum states for a setup that meets the Bragg condition for
three-photon scattering with $n_r=2$ and $n_l=-1$.  Initially the mode
with momentum $\vec p$ is occupied only.  During the interaction with
the laser other modes get occupied as well and after a smooth turn-off
phase the mode with momentum $\vec p+3\hbar \vec k$ is occupied with
probability one.  Generally the end state depends on the total
interaction time $T$.  Numerical simulations show that the occupation
probabilities oscillate in Rabi cycles
\cite{Ahrens:etal:2012:spin_dynamics}.  In the laser field the
probabilities $|c_{n}^{\gamma}(t)|^2$ oscillate on two time scales, a
short scale which is given by the laser period $T_L$ and a large scale
which is given by the Rabi period $T_R$.  The effect of the fast
oscillations disappears when the electron leaves the laser field while
the effect of the slow variation can be measured in the electron's
momentum distribution outside the laser leading to the Kapitza-Dirac
effect, see figure~\ref{fig:KDE} right panel.  Figure~\ref{fig:flip_0_3}
shows the occupation probability $|c_{n}^{\gamma}(t)|^2$ as a function
of the total interaction time $T$ that means when electron is
field-free.  Because experimental parameters have been chosen such that
the three-photon Bragg condition is met the occupation probability
oscillates between $|c_{0}^{\uparrow+}(t)|^2$ and
$|c_{3}^{\uparrow\downarrow+}(t)|^2$.  Note that when the electron has
absorbed three photons from the laser field the electron is in a
superposition of spin-up and spin-down states although the electron beam
enters the laser spin-polarized along the direction of the laser's
electric field component \cite{Ahrens:etal:2012:spin_dynamics}. Thus,
three-photon Kapitza-Dirac scattering induces oscillation of the spin.

\bibliography{computational_qd_Prag_2013}
\bibliographystyle{spiebib}

\end{document}